\documentclass{IEEEoj}
\usepackage{cite}
\usepackage{amsmath,amssymb,amsfonts}
\usepackage{algorithmic}
\usepackage{graphicx,color}
\usepackage{textcomp}
\usepackage{comment}
\def\BibTeX{{\rm B\kern-.05em{\sc i\kern-.025em b}\kern-.08em
    T\kern-.1667em\lower.7ex\hbox{E}\kern-.125emX}}
\AtBeginDocument{\definecolor{ojcolor}{cmyk}{0.93,0.59,0.15,0.02}}
\def\OJlogo{\vspace{-14pt}\includegraphics[height=28pt]{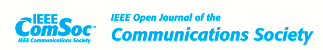}}
\begin{document}
\receiveddate{XX Month, XXXX}
\reviseddate{XX Month, XXXX}
\accepteddate{XX Month, XXXX}
\publisheddate{XX Month, XXXX}
\currentdate{XX Month, XXXX}

\newcommand{\fb}{f_{\mathrm{3dB}}}
\newcommand{\li}{\ell_\infty}
\newcommand{\lc}{\ell_{100}}
\newcommand{\lo}{\ell_{0}}
\newcommand{\fbb}{f^2_{\mathrm{3dB}}}
\newcommand{\lii}{\ell^2_\infty}
\newcommand{\liiii}{\ell^4_\infty}
\newcommand{\lcc}{\ell^2_{100}}
\newcommand{\lcccc}{\ell^4_{100}}
\newcommand{\loo}{\ell^2_{0}}

\newcommand{\fbe}{\tilde{f}_{\mathrm{3dB}}}
\newcommand{\lie}{\tilde{\ell}_\infty}
\newcommand{\lce}{\tilde{\ell}_{100}}
\newcommand{\fbbe}{\tilde{f}^2_{\mathrm{3dB}}}
\newcommand{\liie}{\tilde{\ell}^2_\infty}
\newcommand{\lcce}{\tilde{\ell}^2_{100}}
\newcommand{\intinf}{\int_{-\infty}^{+\infty} }

\newcommand{\Ts}{T_\mathrm{s}}
\newcommand{\Tc}{T_\mathrm{c}}
\newcommand{\Rs}{R_\mathrm{s}}
\newcommand{\Rc}{R_\mathrm{c}}

\newcommand{\bho}{\rho }
\newcommand{\sigmaR}{\frac{\sigma_u^2}{4\pi}}

\newcommand{\real}[1]{\mathcal{R}{\left\{#1\right\}}}
\newcommand{\imag}[1]{\mathcal{I}{\left\{#1\right\}}}

\newtheorem{theorem}{Theorem}
\newtheorem{lemma}{Lemma}

\title{Discrete-time models and performance of phase noise channels}

\author{A. PIEMONTESE$^{\text{1}}$, G. COLAVOLPE$^{\text{1}}$, AND T. ERIKSSON$^{\text{2}}$}
\affil{Department of Engineering and Architecture, University of Parma, Italy, and CNIT research unit}
\affil{Department of Electrical Engineering, Chalmers University of Technology, Gothenburg, Sweden}
\corresp{CORRESPONDING AUTHOR: A. Piemontese (e-mail: amina.piemontese@unipr.it).}
\markboth{Discrete-time models and performance of phase noise channels}{Piemontese\textit{ et al.}}

\begin{abstract}
This paper deals with the phase noise affecting communication systems, where local oscillators are employed to obtain reference signals for carrier and timing synchronizations. 
The most common discrete-time phase noise channel model is analyzed, with the aim to fill the gap between measurements and analytical models. In particular, the power loss and the intersymbol interference due to the presence of phase noise is evaluated with reference to the measurements parameters and to the system bandwidth.
Moreover, the impact on the communication systems' performance of the phase noise originating from the oscillator non idealities is considered, in case of free-running and phase-locked oscillators. The proposed analysis allows to extrapolate useful information about the performance of practical systems by investigating the power spectral density of the oscillator phase noise. An expression for the variance of the residual phase error after tracking, which depends on the main parameters of practical oscillators, is derived, and used to study the dependence of the performance on the symbol rate.
\end{abstract}


\maketitle

\section{Introduction}
\IEEEPARstart{T}{he} phase noise (PN) remains one of the main limiting factors for communication systems. PN due to instabilities of local oscillators, both at the transmitter and at the receiver, can in fact cause a severe performance degradation. Ideally, a local oscillator would produce a sinusoidal signal, whose power spectrum is a delta function at the carrier frequency, but in reality its output power appears also in a band around the desired frequency. 
Since the PN can strongly limit the performance, the study of the PN effects in communication systems has attracted a lot of interest in the literature of the last decades, see, e.g.,~\cite{To98, Ar01,WuBa04,De06,Co14,KhKuPa14,CoCoPi23}.

In~\cite{PiCoEr22}, a PN spectrum model which is very suitable for theoretical calculations is studied. The considered power spectral density (PSD) has the fundamental features typical of the PSD of practical oscillators, i.e., a $-20$~dB/decade slope, a flat part at low frequencies, representing the attenuation of the PN stabilized by means of a PLL, and another flat part at high frequencies. Moreover, it has the Wiener model as a special case. It is similar to other models in the literature~\cite{TuCo05,Mu95,RoKa95}, but it improves them since it allows us to describe the PN in terms of measurement parameters of practical oscillators and facilitates closed-form expressions of distortions and performance. Starting from the PN PSD, the PSD of the phasor can be derived and expressed in terms of the parameters of the PN PSD, that are related to the oscillator measurements. In this paper, the analysis is extended to the general case where the flat part of the PN PSD at high frequencies is not negligile and without resorting to the approximation of low PN as usually done in the literature.

Despite the large number of publications on the subject, there are questions that are not concisely answered yet.
An important point regards a largely used discrete-time channel model, that ignores the power loss and the intersymbol interference (ISI) affecting the received symbols due to the PN. 
The PN process has in fact an infinite bandwidth, and some approximation errors will occur when using a discrete-time model. Further, samples at the output of the matched filter are not a sufficient statistic for detection since the receiver filter is not matched due to the presence of PN. In~\cite{GhKr17}, the effect of filtering on the phase noise is considered by using a multi-sample receiver, in the case of Wiener PN and in the absence of ISI. However, to the authors' knowledge, the quantification of these neglected effects in terms of the measurement parameters is missing in the literature.

Another question that is worth to consider is how can we relate the performance of phase tracking algorithms to the oscillator measurements? It is of particular interest the dependence of the performance on the white oscillator noise floor. This effect has been recently studied in the literature through an experimental approach \cite{ChHeKu17,ChKuGu18}, but a theoretical analysis is still missing. 

In this paper, we give an answer to these questions starting from the aforementioned general analytical model~\cite{PiCoEr22}, which can describe the PSD of the PN of real oscillators.
The approximation errors that the discrete-time representation of the PN suffers from are bounded. For example, to quantify the ISI due to the presence of PN in the case of a free-running oscillator and single carrier modulation, the signal-to-interference (SIR) ratio is derived in closed form as a function of the ratio between the phasor bandwidth and the communication system bandwidth, or, equivalently, as a function of the innovation variance $\sigma_u^2$ of the discrete-time Wiener PN process.
Interestingly, we can find a limit on $\sigma_u$ to have a SIR higher than a given value, e.g., the SIR is higher than $25$~dB if $\sigma_u<0.1$~rad, and this means that for many realistic cases those approximation errors are limited.

Another contribution of this paper is the study of the performance of phase estimators employed at the receiver side. The PSD of the residual phase error after phase tracking is derived and connected to the main measurement parameters of oscillators, to the system bandwidth and to the variance of the additive noise affecting the communication system. The proposed analysis shows that the performance does not depend monotonically on the symbol rate but there exists an optimal value beyond which performance degrades. To the author's knowledge, this is the first time that the optimal symbol rate is found in closed form.

The rest of the paper is organized as follows. 
The adopted statistical model of the PN is described in Section~\ref{s:psd}, while the statistical model of the phasor is given in Section~\ref{s:psd_ph}. Section~\ref{s:dt_channel} introduces the communication system model. The baseband discrete-time PN channel is analyzed in Section~\ref{s:error} and the theory about the performance of phase trackers is derived in Section~\ref{s:phase_tracking}. Numerical results are collected in Section~\ref{s:results}. 

\textit{Notation.} Given a random process $x(t)$, the PSD  and the autocorrelation function are denoted by $S_x(\cdot)$ and  $R_x(\cdot)$, respectively. The expectation with respect to the random variable $y$ is denoted by $E_y\{\cdot\}$.


\section{Statistical model of the phase noise}\label{s:psd}
The single-side-band PN spectrum of many practical oscillators that can be found from measurements is characterized by a $-20$~dB/decade slope~\cite{PiCoEr22} due to integration of white noise inside the oscillator circuitry, and by two flat parts: one at low frequencies, representing the attenuation of the PN stabilized by means of a PLL, the other at high frequencies, modelling the thermal noise at the oscillator output~\cite{De06}. Therefore, the PN PSD can be modeled as
\begin{equation}\label{e:psd_l100}
S_\theta(f)=\frac{10^{10}\lcc}{\fbb+f^2}+\lii\, .
\end{equation}
where $\fb$ is the 3dB bandwidth, and $\lcc$ and $\lii$ are the spectrum levels for $f\!=\!100$~kHz and for high frequencies, respectively. 
The PSD in~(\ref{e:psd_l100}) is obtained by assuming that $\fb\ll100$~kHz\footnote{When the condition $\fb \ll 100$~kHz does not hold, the PSD can simply be expressed as a function of another spectrum level in the $-20$~dB/decade region.} and $\li \ll \lc$. 
The first term in~(\ref{e:psd_l100}) is the PSD of a first-order autoregressive (AR) process~\cite{Me00}, while the flat part at high frequencies dominates when $f> \sqrt{ \frac{10^{10}\lcc}{\lii} -\fb^2}$~\cite{PiCoEr22}. 

In this work, two limiting cases are considered: a free-running oscillator and a PLL-locked oscillator. In particular, the case \mbox{$\fb \rightarrow 0$} is studied, which corresponds to a free-running oscillator. In this case, the PN process is a nonstationary Wiener process with a variance that increases linearly with time; on the other hand, the PN has stationary increments and can be described through the variance of the phase increments. The other case is the case of high $\fb$, i.e., \mbox{$\fb\gg \pi 10^{10} \lcc$},  which can represent a PLL-locked oscillator with 3dB-bandwidth $\fb$. 
This condition is easily met in practical PLL-locked oscillators. In fact, the value of the spectrum at 100 kHz, $\lcc$,  is most often between $-120$ and $-80$ dBc/Hz, therefore the quantity $\pi 10^{10} \lcc$ is in the range $[ 0.03, 300]$~Hz. Since the PLL bandwidth is typically in the order of $1$~kHz or 1-2 decades higher than this value, the condition \mbox{$\fb\gg \pi 10^{10} \lcc$} is normally satisfied for PLL-locked oscillators. 
The range of values of parameters of typical oscillators are reported in Table~\ref{t:parameters}.\footnote{Parameter $\fb$ is given for PLL-locked oscillators.}
\begin{table}[ht]
\centering
\caption{Parameters of typical oscillators.}

\begin{tabular}{|c|c|}
\hline
parameter& range  \\ \hline
\hline
$\fb$   & $[10^3,10^5]$ Hz  \\ \hline
$\lcc$     & $[-120,-80]$ dBc/Hz  \\ \hline
$\lii$     & $<-120$ dBc/Hz \\ \hline
$\pi 10^{10} \lcc$ & $[0.03,300]$ Hz  \\ \hline
\end{tabular}
\label{t:parameters}
\end{table}
\section{Power spectral density of the phasor}\label{s:psd_ph}
Since both PN and phasor PSDs are used by industries and studied in the literature~\cite{PiMa02,BaLe90,GhKr17}, the PSD of the random process of the phasor $\phi(t)\triangleq e^{j\theta(t)}$ is here derived. 
Through simulations, it can be observed that the PSD of the phasor follows the one of the PN at high frequencies. In the literature, see for example~\cite{PiMa02}, this is motivated by using the low-PN approximation, i.e.,
\begin{equation}
\phi(t)\simeq1+j \theta(t)
\end{equation}
when $\theta(t)$ is small. However, this approximation is not always met (quite often it is not met in practice, actually), and more exact expressions are derived below.

The PSD of the random process of the phasor can be derived starting from the general model of the PN in~(\ref{e:psd_l100}). We first consider the case where the flat part of the PN PSD at high frequency is negligible, i.e., $\lii=0$.
Let us define the process $h_\tau(t)$ of the phase increments as
\begin{equation}
h_\tau(t)=\theta(t)-\theta(t-\tau)\,.
\end{equation}
The process $h_\tau(t)$ is a Gaussian random process~\cite{FoVa88} with zero mean and variance 
\begin{align}
\sigma^2_{h_\tau}&=\text{E}\{ (\theta(t)-\theta(t-\tau))^2  \} \\
&= \frac{2\pi 10^{10} \lcc}{\fb}  (1- e^{-2\pi \fb  |\tau|} )  \, ,\label{e:varDelta}
\end{align}
where the following expression of the PN autocorrelation
\begin{equation}\label{e:corr}
R_\theta(\tau)=\frac{\pi 10^{10} \lcc}{\fb}  e^{-2\pi \fb  |\tau|}\, ,
\end{equation} 
obtained by taking the inverse Fourier transform of~(\ref{e:psd_l100}) with $\li=0$, has been used.
The variance (\ref{e:varDelta}) increases with $|\tau|$ up to a ceiling at $\frac{2\pi 10^{10} \lcc}{\fb}$. 
The autocorrelation of the phasor, denoted by $R_\phi(\tau)$, is
\begin{align}
R_\phi(\tau)&=\text{E}\{   e^{j\theta(t)} e^{-j\theta(t-\tau)} \} \\
&= \text{E}\{ e^{jh_\tau(t) }  \} \\ 
&=e^{-\frac{\sigma^2_{h_\tau}}{2}}\label{e:exp_gaus}\, ,
\end{align}
where the fact that $h_\tau(t) $ is a zero-mean Gaussian random variable has been used. 
Replacing~(\ref{e:varDelta}) in~(\ref{e:exp_gaus}), the autocorrelation of the phasor process is obtained as
\begin{equation}\label{e:corr_phasor}
R_\phi(\tau)=e^{- \frac{\pi 10^{10} \lcc}{\fb}   (1- e^{-2\pi \fb  |\tau|} )}\, .
\end{equation}

The PSD of the phasor process is obtained by taking the Fourier transform of the autocorrelation function
\begin{align}\nonumber
S_{\phi}& (f)=e^{-\frac{\pi 10^{10} \lcc}{\fb}}  \delta(f) + \\\nonumber
&e^{-\frac{\pi 10^{10} \lcc}{\fb}}   \int_{-\infty}^{\infty}    (  e^{     \frac{  \pi 10^{10} \lcc e^{-2 \pi \fb |\tau | } }{\fb}   }  -1   ) \mathrm{cos}(2\pi f \tau) \mathrm{d} \tau\, ,
\end{align}
where $\delta(f) $ denotes the Dirac delta function.

Now the two limiting cases \mbox{$\fb \rightarrow 0$} and \mbox{$\fb\gg \pi 10^{10} \lcc$} are considered.
In the first case, using the approximation $e^{-2\pi \fb |t|} \simeq  1 -2\pi \fb |t|$, which is valid when $\fb\rightarrow 0$, the phasor PSD becomes 
\begin{align}\nonumber
S_\phi(f)_{\fb \to 0}  &=  \mathcal{F} \{ e^{-2\pi^2 10^{10}\lcc |\tau| }  \} \\ \label{e:free_phasor_spectrum}
&= \frac{ 10^{10}\lcc}{\pi^2 10^{20} \lc^4  + f^2}\, .
\end{align}
When $\fb\gg \pi 10^{10} \lcc$, the autocorrelation of the phasor can be approximated as
\begin{equation}
R_\phi(\tau)\simeq 1- \frac{\pi 10^{10} \lcc}{\fb}   (1- e^{-2\pi \fb  |\tau|} )
\end{equation}
and the PSD can be computed as
\begin{align}\nonumber
S_\phi(f)_{\text{high}\fb } & = \! \mathcal{F} \Big\{ \!1- \!\frac{\pi 10^{10} \lcc  }{\fb}  \!+ \!\frac{\pi 10^{10}  \lcc  e^{-2\pi \fb |\tau|} }{\fb} \! \Big\}\\ \label{e:pll_phasor_spectrum}
&=\!\Big( 1- \!\frac{\pi 10^{10} \lcc  }{\fb} \Big) \delta(f)   + \frac{  10^{10} \lcc }{\fbb+f^2}\, .
\end{align}
Expressions~(\ref{e:free_phasor_spectrum}) and~(\ref{e:pll_phasor_spectrum}) give the PSD in the case of a free-running oscillator and in the case of a PLL-locked oscillator, respectively. The PSD for the free-running oscillator is a Lorentzian spectrum, with the value at 100~kHz given by $\lcc$ and \mbox{3-dB} bandwidth equal to 
\begin{equation}\label{e:Bphi}
f_{\mathrm{3dB},\phi}=\pi10^{10}\lcc  \, ,
\end{equation}
and overlaps with~(\ref{e:psd_l100}) for $f\gg \pi 10^{10} \lcc$. In the case of the PLL-locked oscillator, the spectrum has a delta function at $f\!=\!0$, and then follows exactly the PN spectrum~(\ref{e:psd_l100}).

The case $\lii\neq0$ is now considered. 
In the following derivation, the bandwidth $B_\theta$ of the white PN is introduced as a help parameter.\footnote{This is necessary to avoid dealing with a white process with infinite bandwidth and power, but the end results are independent of this help parameter.} Ideally, $B_\theta$ can be chosen arbitrarily large. On the other hand, $\lii B_\theta\ll 1$ is assumed in order to bound the error of the used approximations--see~(\ref{e:error_taylor}). In practical oscillators, the value $\lii$ is in the order of \mbox{$-120$}~dBc/Hz or less, therefore, even for large $B_\theta$, the above assumption can be satisfied.
The statistical quantities that refer to the case of $\lii\neq0$ are marked with a circumflex accent.
When the assumption $\lii=0$ is removed, the autocorrelation of the phasor is
\begin{equation}\label{ea:exp_gauss2}
    \hat{R}_\phi(\tau)=R_{\phi}(\tau)  e^{-\lii B_\theta (1-\text{sinc}(\tau B_\theta))}\, ,
\end{equation}
where the autocorrelation of the phasor when $\lii=0$, $R_{\phi}(\tau)$, is given in~(\ref{e:corr_phasor}). 
Using the first order Taylor expansion of the exponential $e^x\simeq 1+x$ in~(\ref{ea:exp_gauss2}), we get
\begin{equation}\label{ea:app}
   e^{-\lii B_\theta (1-\text{sinc}(\tau B_\theta)}= 1-  \lii B_\theta (1-\text{sinc}(\tau B_\theta))+E(\tau)\, ,
\end{equation}
where $E(\tau)$ is the approximation error, upper bounded by the second term of the Taylor expansion~\cite{kline1998calculus},
that is
\begin{equation}\label{e:error_taylor}
   E(\tau)\leq  \liiii B_\theta^2 (1-\text{sinc}(\tau B_\theta))^2 /2\leq \liiii B_\theta^2\, .
\end{equation}
As stated, this error is low for a large range of values of $B_\theta$. If, for example, we consider $\lii=-120$~dB, we have an error smaller than $10^{-4}$ for $B_\theta<10^{10}$. For this reason, we neglect $E(\tau)$ in the following.
Using~(\ref{ea:app}) in~(\ref{ea:exp_gauss2}), we get
\begin{equation}
\hat{R}_\phi(\tau)=R_{\phi}(\tau)  (1 -\lii B_\theta + \lii B_\theta\text{sinc}(\tau B_\theta))\, ,
\end{equation}
and the PSD of the phasor is obtained by taking the Fourier transform of the above autocorrelation, i.e.,
\begin{equation}\label{e:s2}
    \hat{S}_\phi(f)=(1 -\lii B_\theta) S_{\phi}(f)    + S_{\phi}(f) \otimes \lii \Pi\Big(\frac{f} {B_\theta}\Big)\, ,
\end{equation}
where symbol $\otimes$ denotes convolution.
In~(\ref{e:s2}), the term \mbox{$1 -\lii B_\theta$} that multiplies $S_{\phi}(f)$ makes the power of the phasor unitary, while the effect of the flat PN, dominating at high frequencies, is represented by the second term, which, in the case of free-running oscillator, is
\begin{align}\nonumber
    S_{\phi}(f) \otimes  \lii \Pi\Big(\frac{f} {B_\theta}&\Big)= \frac{\lii }{\pi} \Big[ \text{tan}^{-1} \Big(\frac{1}{\pi 10^{10}\lcc} \Big(f+\frac{B_\theta}{2}  \Big)\Big)\\
    &-\text{tan}^{-1} \Big(\frac{1}{\pi 10^{10}\lcc} \Big(f-\frac{B_\theta}{2}  \Big)\Big)   \Big]\, .\label{e:flat}
\end{align}
\begin{figure}
\centering
\includegraphics[width=\columnwidth]{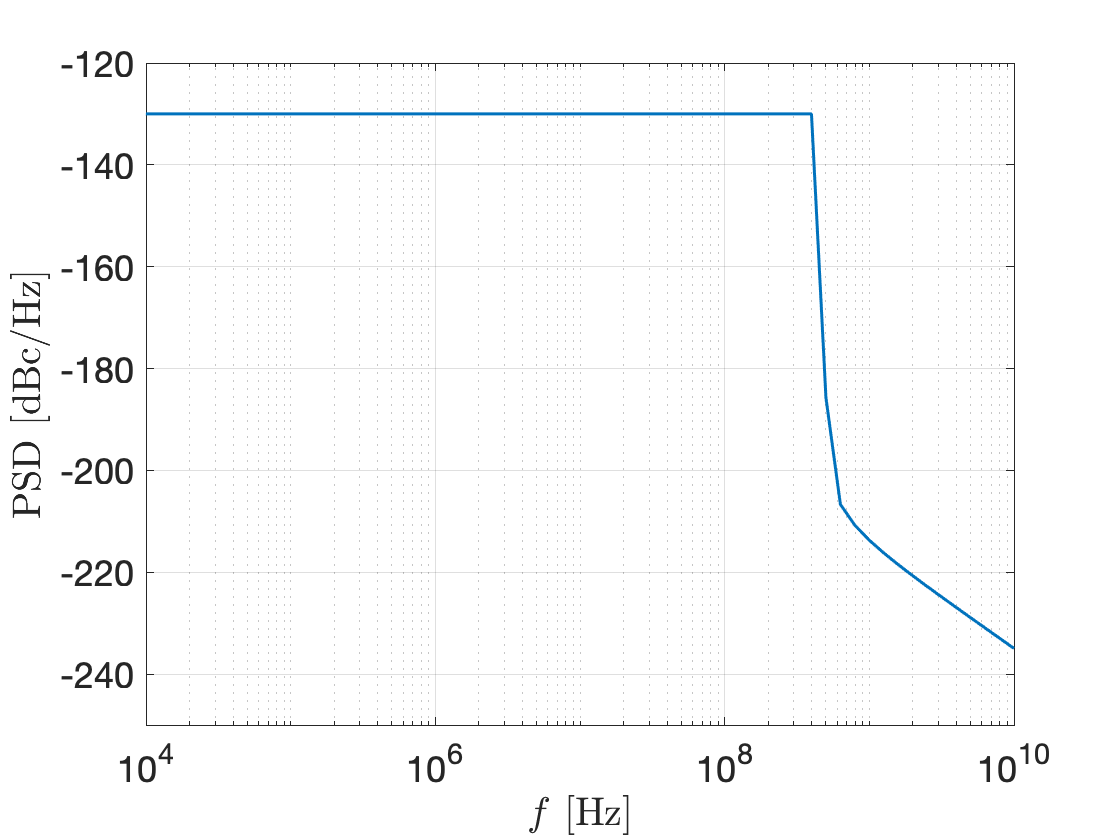}
\caption{Flat part of the phasor power spectral density given in~(\ref{e:flat}), for $B_\theta=10^9$, $\lii=-130$~dBc/Hz and $\lcc=-90$~dBc/Hz.}\label{f:flat}
\end{figure}
This term is constant and equal to $\lii$ for $|f|\ll B_\theta$ and equal to zero for $|f|\gg B_\theta$. It is shown in Figure~\ref{f:flat} for $B_\theta=10^9$, $\lii=-130$~dBc/Hz and $\lcc=-90$~dBc/Hz.

Using the approximation $\lii B_\theta\ll 1 $, and assuming $1/\Ts\ll B_\theta$, the PSD of the phasor in the communication band can be written with good approximation as
\begin{equation}\label{e:psd_plus_flat}
        \hat{S}_\phi(f)=S_{\phi}(f)+ \lii\, .
\end{equation}

\section{Application to communications}\label{s:dt_channel}
A communication system is considered, where linearly modulated symbols $\{x_n\}$ are transmitted through the channel. At the receiver, after down-conversion, the received signal is 
\begin{equation}\label{e:r1}
r(t)=\sum_n x_n p(t-n \Ts) e^{j \theta_\text{T}(t)} e^{j \theta_\text{R}(t)}+ v(t)e^{j \theta_\text{R}(t)}\, ,
\end{equation}
where $\Ts$ is the symbol time, $p(t)$ is the shaping pulse with unitary energy satisfying the Nyquist criterion, $v(t)$ is the complex  additive white Gaussian noise (AWGN) with one-sided PSD $N_0$, and $\theta_\text{T}(t)$ and $\theta_\text{R}(t)$ are the PN processes that arise from local oscillator instabilities at the transmitter and at the receiver, respectively (see Section~\ref{s:psd}). 
The model~(\ref{e:r1}) can be rewritten as
\begin{equation}\label{e:r}
r(t)= \sum_n x_n p(t-n \Ts) e^{j \theta(t)} + w(t)\, ,
\end{equation}
where $w(t)$ is an AWGN with same statistics of $v(t)$ and $\theta(t)$ is the sum of transmitter and receiver PN processes. The PSD of $\theta(t)$ is the sum of the PSDs of $\theta_\text{T}(t)$ and $\theta_\text{R}(t)$, where the two PN processes are independent. In many practical cases, receiver and transmitter oscillators have quite different performance, and hence only the dominant PN can be considered, i.e., the PN of the oscillator adopted in consumer grade equipment. In other cases, they are similar and $\theta(t)$ has the sum level of both PNs. Hence, the PSD model described in Section~\ref{s:psd} is applicable in both cases.

\begin{figure}
\centering
\includegraphics[width=\columnwidth]{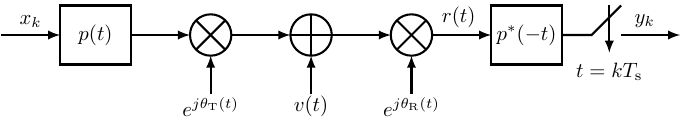}
\caption{Block diagram of the system model.} \label{f:blockdiagram}
\end{figure}

At the receiver side, a filter matched to the shaping pulse $p(t)$ is employed, followed by a sampler at symbol time. The received samples represent an approximate sufficient statistic and are given by
\begin{equation}\label{e:rec}
y_k=r(t)\otimes p^*(-t)_{|_{t=k \Ts}}\, .
\end{equation}
The considered system model is shown in Figure~\ref{f:blockdiagram}.
A commonly studied discrete-time channel model is \begin{equation}\label{e:discretetime}
z_k= x_k e^{j \theta_k}  + w_k \, ,
\end{equation}
where $\theta_k$ is the PN sample obtained as $\theta_k =\theta(k \Ts)$ and $w_k$ is the sample obtained by filtering and sampling the AWGN signal $w(t)$. 
The discrete-time model for the PN with PSD~(\ref{e:psd_l100}) is
\begin{equation}\label{e:theta}
    \theta_k=\theta^c_k+\theta^f_k\, 
\end{equation}
where $\theta^c_k$ and $\theta^f_k$ are independent processes, the former inducing the $-20$~dB/decade slope in the PN PSD, which dominates close-to-carrier, the latter inducing the flat part of the PSD, that dominates at high frequencies. Random variables  $\{\theta^f_k\}$ are independent and identically distributed (i.i.d.) zero-mean Gaussian variables with variance $\sigma^2_f=\lii/\Ts$, while $\theta^c_k$ is a discrete-time AR random process 
\begin{equation}\label{e:dt model}
\theta^c_k=a\theta^c_{k-1}+u_k\, ,
\end{equation}
where $u_k$ are i.i.d. zero-mean Gaussian variables with variance $\sigma^2_u$, and parameters $a$ and $\sigma^2_u$ are
\begin{equation}\label{e:dt_parameters}
a=e^{-2\pi \fb \Ts}\,, \quad
\sigma^2_u=\frac{\pi 10^{10} \lcc}{\fb} (1-e^{-4\pi \fb \Ts})\, .
\end{equation}
The case of a free-running oscillator implies that $\fb\to 0$. In this case, the parameter $a$ is equal to 1 and the model~(\ref{e:dt model}) reduces to the Wiener (random-walk) model. The Wiener process is nonstationary, but the process of the phase increment $u_k$ is stationary and the PN can be described through the variance $\sigma^2_u$. We can compute the variance of the phase increments in~(\ref{e:dt_parameters}) by taking the limit for $\fb\rightarrow 0$ and we obtain
\begin{align}
\sigma^2_u&= 4 \pi^2 10^{10}\lcc \Ts   \label{e:innovation variance} \\
&=4\pi\bho \, , \label{e:innovation variance rho}
\end{align}
where $\bho\triangleq f_{\mathrm{3dB},\phi} \,\Ts$ represents the ratio between the phasor \mbox{3-dB} bandwidth and the signal bandwidth and will be called \textit{relative bandwidth parameter} in the following.

Model~(\ref{e:discretetime}) neglects the ISI and the power loss due to the presence of PN. In fact, the PN causes an enlargement of the signal bandwidth, therefore the filter matched to the shaping pulse $p(t)$ cuts some spectral components of the signal, resulting in a power loss. Moreover, the received samples are affected by ISI due to mismatched filters. These effects are quantified in Section~\ref{s:error}.


\section{Error analysis of the discrete-time channel}\label{s:error}
Here, the accuracy of the communication model~(\ref{e:discretetime}) is evaluated. Thanks to the statistical model described in Section~\ref{s:psd_ph}, the results of this analysis can be connected to the main PN parameters of practical oscillators. 
The case of a free-running oscillator with $\lii=0$ is studied, for which the PSD of the phasor is given by~(\ref{e:free_phasor_spectrum}). The case $\lii\neq 0$ is considered at the end of the section.

The following mean square error (MSE) is computed and related to the parameters of the PN
\begin{equation}\label{e:eta1}
\eta=E_{x,\theta,w}\Big \{ |z_k-y_k|^2\Big \}\, ,
\end{equation}
where the expectation is done with respect to the transmitted symbols, the PN and the AWGN. In~(\ref{e:eta1}), $z_k$ is the received sample of the simplified model~(\ref{e:discretetime}) ignoring ISI and power loss, while $y_k$ is the received sample of the more accurate model~(\ref{e:rec}). In order to describe the effect of the PN and matched filtering on the received samples, we define the following linear time-varying discrete filter
\begin{equation}\label{e:g}
    g_{n,m}=p(t)e^{j\theta(t+m T_s)}\otimes p^*(-t)|_{{t=n T_s}}\,.
\end{equation}
In the absence of PN and for a Nyquist shaping pulse, \mbox{$g_{0,m}=1$} and $g_{n,m}=0$ for $n\neq0$.
\begin{lemma}\label{l:eta} \textit{Consider the system model described in~(\ref{e:r}), where the phasor $e^{j\theta(t)}$ has PSD given in~(\ref{e:free_phasor_spectrum}). Let $e^{j \theta_k}$ be the sample $e^{j \theta(k T_s)}$. The MSE in~(\ref{e:eta1}) between $z_k$ defined in~(\ref{e:discretetime}) and $y_k$ defined in~(\ref{e:rec}) is given by
\begin{equation}\label{e:eta}
\eta=  \eta_\mathrm{D} + \eta_\mathrm{ISI}  \, ,
\end{equation}
where
\begin{align}\label{e:etad}
\eta_\mathrm{D}=&  E_{\theta}\Big \{ |  e^{j \theta_k} -  g_{0,k}|^2\Big \}    \, , \\ \label{e:etaisi}
\eta_\mathrm{ISI}=&   \sum_{m\neq 0}  E_{\theta}\Big \{ | g_{m,k-m}  |^2\Big \}    \, ,  & 
\end{align}
and the terms $g_{n,m}$ are given in~(\ref{e:g}).}
\end{lemma}
\begin{IEEEproof}
By expanding~(\ref{e:rec}), we obtain
\begin{align}\label{e:rec_k}
 y_k& =\sum_m x_{k-m}  g_{m,k-m}    +w_k\\
\label{e:y_k}
&= x_k g_{0,k} + \sum_{m\neq 0} x_{k-m} g_{m,k-m}   +w_k \, ,
\end{align}
where $g_{0,k}$ is connected to a power loss of the desired output $x_k$, and the summation represents ISI. 
Using (\ref{e:discretetime}) and (\ref{e:y_k}) in (\ref{e:eta1}), we obtain
\begin{equation}\nonumber
\eta=E_{x,\theta}\Big \{ |  x_k e^{j \theta_k} - x_k g_{0,k} - \sum_{m\neq 0} x_{k-m} g_{m,k-m}      |^2\Big \}  \, .
\end{equation}
Assuming that the transmitted symbols are zero mean, i.i.d., and with unitary energy, $\eta$ becomes the sum of two errors, one due to the approximation of $g_{0,k}$ with the phasor $e^{j \theta_k}$, the other due to the ISI, that is
\begin{equation}\nonumber
\eta= E_{\theta}\Big \{ |  e^{j \theta_k} -  g_{0,k}|^2\Big \}  +\sum_{m\neq 0} \!\! E_{\theta}\Big \{ | g_{m,k-m}  |^2\Big \} .
\end{equation}
As expected, in the absence of PN and for a Nyquist shaping pulse, $g_{0,k}=1$ and $g_{m,k}=0$ for $m\neq0$.
\end{IEEEproof}
For convenience, the following terms are defined
\begin{equation}\label{e:gammal}
    \gamma_m= E_{\theta}  \{  | g_{m,k-m} |^2  \}\, ,
\end{equation}
which, using the definition of $g_{n,m}$ in Lemma~\ref{l:eta}, can be written as
\begin{align}\nonumber
\gamma_m& =  \intinf \intinf   p(t)p^*(t_1)p^*(t- m T_s) p( t_1- m T_s) \\ &\;\;\;\;\cdot R_\phi(t_1-t) d t d t_1, \label{e:gamma_l}
\end{align}
where $R_\phi(\tau)$ is the autocorrelation of the phasor $\phi(t)=e^{j \theta(t)}$. Equation~(\ref{e:gamma_l}) shows that the terms $\gamma_m $ are independent of $k$. When a detector that is not able to account for the ISI is employed, the SIR can be expressed as a function of the terms $\gamma_m$.
Using the assumption of zero mean symbols
and considering that the symbols are independent, with unitary energy, we have
\begin{equation}\label{e:sir}
\text{SIR}=\frac{\gamma_0}{ \sum_{m\neq 0} \gamma_m}\, .
\end{equation}

The following theorem gives the expression of the MSEs in Lemma~\ref{l:eta} as a function of the PN parameters and of the system bandwidth. The theory is derived by assuming that the filter $p(t)$ is a root-raised cosine (RRC) shaping pulse with roll-off factor zero, i.e., a normalized sinc function. Other roll-off factors will be considered later in this paper.
\begin{theorem}\label{thm:eta}
\textit{Consider the system model described in~(\ref{e:r}), where $p(t)$ is the normalized sinc function with unitary energy, and the phasor $e^{j\theta(t)}$ has PSD given in~(\ref{e:free_phasor_spectrum}), with bandwidth $f_{\mathrm{3dB},\phi}$~(\ref{e:Bphi}). The MSEs $\eta$, defined in~(\ref{e:eta1}), $\eta_\mathrm{D}$ and $\eta_\mathrm{ISI}$, defined in Lemma~\ref{l:eta}, are given by
\begin{equation}\label{e:etarho}
    \eta=1 +\frac{2  }{\pi}\Big[ \frac{\bho}{2} \mathrm{log}\Big(1+\frac{1}{\bho^2}\Big) -\mathrm{tan}^{-1}   \Big( \frac{1}{\bho} \Big)  \Big]\, ,
\end{equation} 
\begin{equation}\label{e:etaDfinal}
\eta_\mathrm{D}=1 +\frac{2 }{\pi}  \Big[  \bho -(1+\bho^2)     \mathrm{tan}^{-1}   \Big( \frac{1}{\bho} \Big) \Big]\,,
\end{equation} 
\begin{equation}\label{e:etaisifinal}
\eta_\mathrm{ISI}= \frac{2  }{\pi}\Big[ \bho^2 \mathrm{tan}^{-1}   \Big( \frac{1}{\bho} \Big)  +  \frac{\bho}{2} \mathrm{log}\Big(1+\frac{1}{\bho^2}\Big)       -\bho\Big]\, ,
\end{equation} 
where $\bho=f_{\mathrm{3dB},\phi} \,\Ts$.}
\end{theorem}
\begin{IEEEproof}
The MSE between the phasor $e^{j \theta_k}$ and $g_{0,k}$ is given in~(\ref{e:etad}). Expanding~(\ref{e:etad}), we obtain
\begin{equation}\label{e:etaD}
\eta_\text{D}=1 + \gamma_0 - 2 \Re\Big( E_{\theta}\Big \{  e^{-j \theta_k} g_{0,k} \Big\} \Big )\,,
\end{equation}
where $\gamma_0$ is given in~(\ref{e:gamma_l}) with $m=0$. 
First the computation of $\gamma_0$ is considered. The expression of the pulse $p(t)$,
\begin{equation}\label{e:sinc}
p(t)=\frac{1}{\sqrt{T_s}} \text{sinc}\Big(\frac{t}{T_s}  \Big)\, ,
\end{equation}
is substituted in~(\ref{e:gamma_l}) with $m=0$,
\begin{equation}
\gamma_0= \frac{1}{{T_s}^2 }\!\! \!  \intinf  \! \!\! \intinf  \!\!\! \text{sinc}^2\Big(\frac{t}{T_s}  \Big) \text{sinc}^2\Big(\frac{t_1}{T_s}  \Big)  \!R_\phi(t_1\!\!-\!\!t) d t d t_1, \label{e:gamma0f}
\end{equation}
then $R_\phi(t_1\! -\!t) $ is replaced with 
\begin{equation}\nonumber
R_\phi(t_1\!-\!t) \!\!= \!\!  \mathcal{F}^{-1} \{ S_\phi (f) e^{-j 2\pi t f} \}\!\! = \!\!\!\! \intinf  \! \!\!  \!S_\phi (f) e^{j 2\pi t_1 f}\! e^{-j 2\pi t f} \!df 
\end{equation}
and the integrals in~(\ref{e:gamma0f}) are rearranged in the following way
\begin{align}\nonumber
\gamma_0=& \frac{1}{{T_s}^2 } \intinf \Big(  \intinf  \text{sinc}^2\Big(\frac{t}{T_s}  \Big)  e^{-j 2\pi t f}    \!d t  \Big) \\\label{e:integral}
&\Big(  \intinf  \text{sinc}^2\Big(\frac{t_1}{T_s}  \Big) e^{j 2\pi t_1 f} d t_1   \Big)  S_\phi (f)   d f\,.
\end{align}
The two internal integrals in $t$ and $t_1$ equal the Fourier transform of $\text{sinc}^2$, that is
\begin{equation}
\intinf \text{sinc}^2\Big(\frac{t}{T_s}  \Big)  e^{-j 2\pi t f}  d t= T_s \Lambda(f T_s),
\end{equation}
where $\Lambda(x)$ is the triangular function, i.e., $\Lambda(x)=1-| x|$, for $-1<x<1$ and zero otherwise. Replacing the expression of $S_\phi (f)$ given in~(\ref{e:free_phasor_spectrum}) in~(\ref{e:integral}), the following expression is obtained
\begin{equation}\label{e:gamma0}
\gamma_0=\frac{2}{\pi}  \Big[   \text{tan}^{-1}   \Big( \frac{1}{\bho} \Big) \Big[ 1 -\bho^2 \Big] -\bho \text{log}\Big(1+\frac{1}{\bho^2}\Big) +\bho \Big]\, ,
\end{equation}
where $\bho=\pi 10^{10} \lcc T_s= f_{\mathrm{3dB},\phi} \,\Ts$. The last term in~(\ref{e:etaD}) can be computed as
\begin{align}\nonumber
E_{\theta}\Big \{ \! e^{-j \theta_k} g_{0,k} \! \Big\}\!\! &=\frac{1}{T_s} \!\!\intinf \!\!\!\! \text{sinc}^2\Big(\frac{t}{T_s}\Big)  R_\phi(t) dt \\  
\label{e:corr_g}
&= \frac{2}{\pi}  \Big[   \text{tan}^{-1}   \Big( \frac{1}{\bho} \Big)-\frac{\bho}{2} \text{log}\Big(1+\frac{1}{\bho^2}\Big)  \Big]\, .
\end{align}
Using~(\ref{e:gamma0}) and~(\ref{e:corr_g}) in~(\ref{e:etaD}), equation~(\ref{e:etaDfinal}) is obtained, concluding the proof for $\eta_\text{D}$.

Now the computation of $\eta_\text{ISI}$ is considered. Using the definition of $\gamma_m$ in~(\ref{e:gammal}), we can write 
\begin{equation}\label{e:etaisi1}
\eta_\text{ISI}=\sum_{m\neq 0} \gamma_m\, .
\end{equation}
In the following, the above summation is computed over all $m$, then $\gamma_0$ in~(\ref{e:gamma0}) is subtracted to the sum to obtain $\eta_\text{ISI}$. Signal $r'(t)$ is defined as the received signal without AWGN, i.e.,
\begin{equation}\label{e:r'}
r'(t)= \sum_n x_n p(t-n \Ts) e^{j \theta(t)}\, .
\end{equation}
Similarly,  $y'(t)$ and $y'_k$ are the signal after matched filtering and the samples after sampling, respectively, without AWGN. The process $r'(t)$ is cycle-stationary, and can be made wide-sense stationary with the classical approach of introducing a random delay, uniformly distributed in $[-T_s,T_s]$. The power of the process $y'(t)$ can be computed by integrating its PSD, that is
\begin{align}\label{e:Sr1}
&P_{y'}= \int_{-\infty}^{+\infty} S_{r'} (f) |P(f)|^2 d f\\
\label{e:Sr}
&=\! \frac{T_s}{\pi} \!\!\int_{-\frac{1}{2T_s}}^{+\frac{1}{2T_s}} \!\!\text{tan}^{-1}\!\Big(\frac{f+1/2T_s}{\pi 10^{10} \lcc}\Big) \!\!-\!\!\text{tan}^{-1}\!\Big(\frac{f-1/2T_s}{\pi 10^{10} \lcc }\Big) d f\\
& =\frac{2}{\pi} \Big[ \text{tan}^{-1}\Big (\!\frac{1}{\bho}    \Big) - \frac{\bho}{2} \text{log}\Big(  1 + \frac{1}{\bho^2}    \Big)  \Big]\label{e:Py}\, .
\end{align}
In~(\ref{e:Sr1}), $P(f)$ is the Fourier transform of $p(t)$. In~(\ref{e:Sr}), the fact that 
the PSD of the received signal is given by the convolution of the PSD of the information carrying  signal and that of the phasor $e^{j\theta(t)}$ has been used. Since the process $y'(t)$ is bandlimited with bandwidth $1/T_s$, the following relationship holds 
\begin{equation}\label{e:Pcd}
P_{y'}= E_{x,\theta}  \Big\{ |y'_k|^2    \Big \} = \sum_m \gamma_m \, .
\end{equation}
By using (\ref{e:Py}) and (\ref{e:Pcd}), we obtain
\begin{equation}\label{e:sumgamma}
\sum_m  \gamma_m = \frac{2}{\pi} \Big[ \text{tan}^{-1}\Big (\!\frac{1}{\bho}    \Big) - \frac{\bho}{2} \text{log}\Big(  1 + \frac{1}{\bho^2}    \Big)  \Big] \, ,
\end{equation}
and from~(\ref{e:etaisi1}) and~(\ref{e:gamma0}) we get~(\ref{e:etaisifinal}). Finally, equation~(\ref{e:eta}) is obtained by using~Lemma~\ref{l:eta}.
\end{IEEEproof}

The expressions above show that the error made when using the channel model~(\ref{e:discretetime}) depends only on the relative bandwidth parameter. 
The SIR can be expressed as a function of $\bho$: starting from~(\ref{e:sir}), the expression of $\gamma_0$ in~(\ref{e:gamma0}) is used in the numerator, while the denominator is $\eta_\text{ISI}$, i.e.,
\begin{equation}\label{e:sir_1}
\text{SIR}=\frac{   \text{tan}^{-1}   \Big( \frac{1}{\bho} \Big) \Big[ 1 -\bho^2 \Big] -\bho \text{log}\Big(1+\frac{1}{\bho^2}\Big) +\bho}{ \bho^2 \mathrm{tan}^{-1}   \Big( \frac{1}{\bho} \Big)  +  \frac{\bho}{2} \mathrm{log}\Big(1+\frac{1}{\bho^2}\Big)       -\bho}\, .
\end{equation}
The SIR can be also expressed as a function of the phase increment standard deviation by using equation~(\ref{e:innovation variance rho}) which connects $\bho$ with $\sigma_u$, i.e.,
\begin{equation}\label{e:sir_2}
\text{SIR}\!=\!\frac{   \text{tan}^{-1} \!  \Big( \frac{4 \pi}{\sigma_u^2} \Big)\! \Big[ 1\!\! -\!\!\frac{\sigma_u^4}{16\pi^2} \Big] \!-\!\frac{\sigma_u^2}{4\pi} \text{log}\Big(1+\frac{16\pi^2}{\sigma_u^4}\Big) + \frac{\sigma_u^2}{4\pi} }{ \frac{\sigma_u^4}{16\pi^2} \mathrm{tan}^{-1}   \Big( \frac{4 \pi}{\sigma_u^2}   \Big)  +  \frac{\sigma_u^2}{8 \pi}  \mathrm{log}\Big(1+\frac{16\pi^2}{\sigma_u^4}\Big)       -\frac{\sigma_u^2}{4 \pi} }\, .
\end{equation}
From the above equation, a limit on $\sigma_u$ can be computed to have a SIR higher than a given value. For example, a SIR higher than $25$~dB requires $\sigma_u<0.1$ rad. 
The extension to the case where $\li$ is not negligible is performed by using equation~(\ref{e:psd_plus_flat}) and we obtain 
\begin{equation}
  \text{SIR}=\frac{\gamma_0 +\frac{2}{3}\frac{\lii}{T_s}}{\sum_{m\neq 0} \gamma_m + \frac{1}{3}\frac{\lii}{T_s}}\, .  
\end{equation}

\section{Optimal phase tracking theory}\label{s:phase_tracking}
In this section, we show how to apply the optimal estimation theory in the context of phase tracking. 
We consider the case where the power loss and the ISI studied in the previous section are negligible, i.e., $y_k\simeq z_k$. This hypothesis is removed in Section~\ref{s:results}, where the PN is generated according to the general model (\ref{e:rec}). 
\subsection{Auxiliary channel models}
The estimator design is based on auxiliary channel models, whose accuracy depends on the SNR with opposite trend. Interestingly, the two models lead to the same expression of the residual PN error after tracking. 
\subsubsection{Low PN approximation}
Using the low PN approximation, $e^{j\theta_k}\simeq 1+j\theta_k$, we get the following relationship
\begin{equation}
    \imag{ y_k/x_k}\simeq\theta_k + \imag{w_k/x_k}
\end{equation}
We define $d^{(1)}_k=\imag{y_k/x_k}$ and $n^{(1)}_k=\imag{w_k/x_k}$ and obtain
\begin{equation}\label{lowpn}
    d^{(1)}_k\simeq\theta_k +n^{(1)}_k
\end{equation}
where $n^{(1)}_k$ has variance
\begin{equation}
    \text{var}(n^{(1)}_k)=\frac{1}{|x_k|^2}\frac{\sigma^2_w}{2}\, .
\end{equation}
\subsubsection{High SNR approximation}
The modulus of the observable $y_k$ is
\begin{align}
	|y_k| = & \left|x_k e^{j\theta_k} + w_k\right| \\
		  = & \left||x_k| e^{j\angle x_k} e^{j\theta_k} + w_k' e^{j\angle x_k} e^{j\theta_k}\right| \\
		  = & \left||x_k| + w_k'\right| 
\end{align}
where $w_k'$ is statistically equivalent to $w_k$. From the above equation, it can be seen that the modulus of the observable cannot be used for the phase estimation. On the other hand, the phase of $y_k$ can be expressed as 
\begin{align}
	\angle y_k = & \angle \left( x_k e^{j\theta_k} + w_k \right) \\ 
		       = & \angle \left( |x_k| e^{j\angle x_k} e^{j\theta_k} + w_k' e^{j\angle x_k} e^{j\theta_k} \right) \\
		       = & \angle x_k + \theta_k + \angle\left( |x_k| + w_k' \right) \\
		       = & \angle x_k + \theta_k + \tan^{-1} \frac{\imag{w_k'}}{|x_k| + \real{w_k'}}\label{e:disctremodel}
\end{align}
In the asymptotic case of high SNR, we have
\begin{equation}
\tan^{-1} \frac{\imag{w_k'}}{|x_k| + \real{w_k'}}\simeq \frac{\imag{w_k'}}{|x_k| }\,. 
\end{equation}
We define the following auxiliary variable
\begin{equation}
d^{(2)}_k= (\angle y_k -\angle x_k)_{\!\!\!\!\!\!\mod\! 2\pi}\,,\label{e:vk}
\end{equation}
where the operator $(\cdot)_{\!\!\!\mod\! 2\pi}$ represents a wrapping of the phase in the interval $(0,2\pi]$, and we obtain the second auxiliary channel model
\begin{equation}\label{hsnr}
d^{(2)}_k\simeq \theta_k +n^{(2)}_k\, ,
\end{equation}
where $n^{(2)}_k=\imag{w_k'}/|x_k| $ has variance 
\begin{equation}
    \text{var}(n^{(2)}_k)=\frac{1}{|x_k|^2}\frac{\sigma^2_w}{2}\,.
\end{equation}

It is important to note that the two above approximations $d^{(1)}_k$ and $d^{(2)}_k$ lead to the same channel model, being the statistics of the noise in~(\ref{hsnr}) and that in~(\ref{lowpn}) the same. Therefore, they will lead to the same expression of the optimal estimation filter. 

\begin{figure}
\centering
\includegraphics[width=\columnwidth]{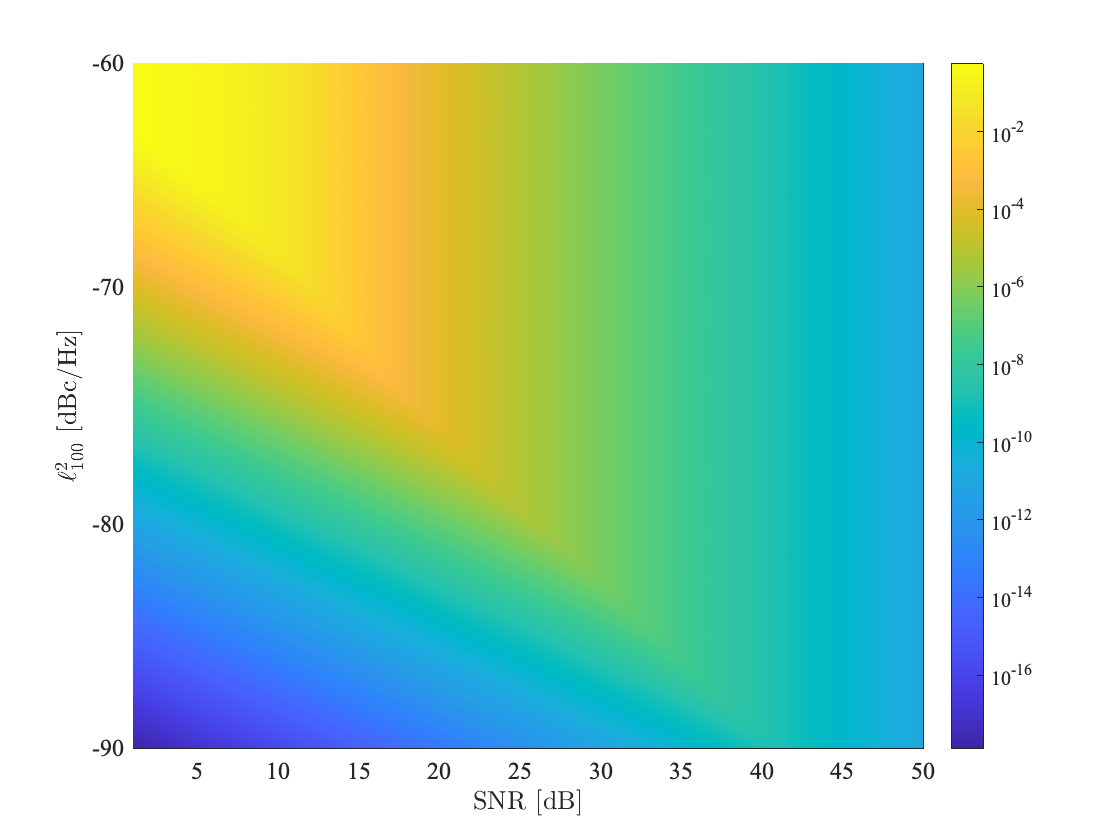}
\caption{Normalized variance of the approximation error for the auxiliary channel models on which the design of the phase estimator is based.}\label{f:error_dt_model}
\end{figure}

In the following, the approximation error of the above models is studied. The exact model can be written as
\begin{equation}
D_k=d^{(i)}_k+e^{(i)}_k,
\end{equation}
for $i=1,2$, where we defined the errors as
$$
e^{(1)}_k=\theta_k-\imag{e^{j\theta_k}}
$$

$$
e^{(2)}_k=\tan^{-1} \frac{\imag{w_k'}}{|x_k| + \real{w_k'}}- \frac{\imag{w_k'}}{|x_k| }\,,
$$
and compute
\begin{equation}\label{e:etak}
\eta^{(i)}_k=\text{var}(e^{(i)}_k)/\text{var}(n^{(i)}_k)
\end{equation}
The variance of the noise sequence in the above equation is time-varying since it depends on $|x_k|$, and for this reason we average it over the transmitted symbols. In Figure~\ref{f:error_dt_model}, we report the minumum for $i=1,2$ of the normalized variance of the error, $\min\{\eta^{(1)}_k,\eta^{(2)}_k\}$ as a function of the SNR and $\lcc$.
In the case of the high SNR approximation, the variance is independent of the PN parameters, while for the low PN approximation the PN sequence has been generated according to the following parameters: $\Rs=1$~MHz, $\fb=10$~kHz and $\lii=-120$~dBc/Hz. From the figure, it can be seen that for all points the approximation error is very small, except for a region, the light yellow one, where neither of the two approximations works, being the PN particularly strong and the SNR low. It is worth noticing that a normalized error of $10^{-2}$ is already very low and corresponds to a system dominated by AWGN.

In the rest of the paper, we will denote by $d_k$ and $n_k$ the samples that we use for the phase estimation and the noise, respectively, leaving out the apex for simplicity.

\subsection{Residual error after phase tracking}\label{s:optimal_tracker}
In this section, the optimal phase tracker is defined, based on the model above. Due to its nature, the noise $\theta_k^f$ in~(\ref{e:theta}) cannot be tracked and hence it will be treated as additional AWGN for the derivation of the phase tracker and its variance added to the variance of the residual error.
The optimal filter is derived in the hypothesis that the transmitted symbols are known (data-aided estimator). To make the analysis feasible, the focus is on a time-invariant estimator, by which the estimate of the PN samples, $\hat{\theta}_k$, can be obtained as the output of a linear time-invariant noncasual filter, i.e., the noncausal Wiener filter, with input $d_k$ and impulse response $q_i$~\cite{Pr96}, that is
\begin{equation}\label{e:filter}
\hat{\theta}_k=\sum_{i=-\infty}^{\infty}  q_i d_{k-i}\, .
\end{equation}
The above assumptions may seem strong, but we will show in Section~\ref{s:results} that the derived closed-form expressions can predict the behavior of practical PN estimators.

We now consider the noise affecting the observations $d_k$.
In each time instant, the variance of $n_k$ depends on the modulus of the transmitted symbol, $|x_k|$, therefore the sequence is non-stationary, and the optimal MMSE estimator for this case is a time-varying filter. In order to find the optimal time-invariant filter, we assume that the noise affecting the observations is $n'_k$ with variance
\begin{equation}\label{e:noise}
\sigma_{n'}^2= \frac{\lii}{\Ts} + \beta \frac{\sigma^2_w}{2}\, ,
\end{equation}
and PSD 
\begin{equation}\label{e:PSDnoise}
S_{n'}(f)=\lii+\frac{\beta \sigma_w^2\Ts }{2}\, ,
\end{equation}
where we included the contribution of the noise $\theta_k^f$. The parameter $\beta$ is greater than or equal to $1$. Particularly, two cases can be identified: one with $\beta=1$, the other with
\begin{equation}\label{e:beta}
\beta = \text{E}_x \Big\{ {\frac{1}{|x_k|^2}}  \Big\}\, .
\end{equation}
The case $\beta=1$ is representative of the phase estimator that is able to exploit the knowledge of the transmitted symbols for determining the variance of the additive noise $n_k$, such as the Kalman smoother that will be considered in Section~\ref{s:results}.

Finally, the system model that we use for the design of the optimal filter for PN tracking is
\begin{equation}\label{e:model}
d_k= \theta^c_k+n'_k\,.
\end{equation}
Assuming the model~(\ref{e:model}), the optimal time-invariant MMSE filter, i.e., the noncausal Wiener filter, for the estimation of the phase according to~(\ref{e:filter}) has Z-transform~\cite{Pr96}
\begin{equation}
\mathbf{Q}(z)=\frac{\mathbf{S}_{\theta^c}(z)}{\mathbf{S}_{\theta^c}(z)+\mathbf{S}_{n'}(z)}\, ,
\end{equation}
where $\mathbf{S}_{\theta^c}(z)$ and $\mathbf{S}_{n'}(z)$ are the Z-transforms of the autocorrelation of the sequence $\theta^c_k$ and $n'_k$, respectively. 

\begin{figure}
\centering
\includegraphics[width=\columnwidth]{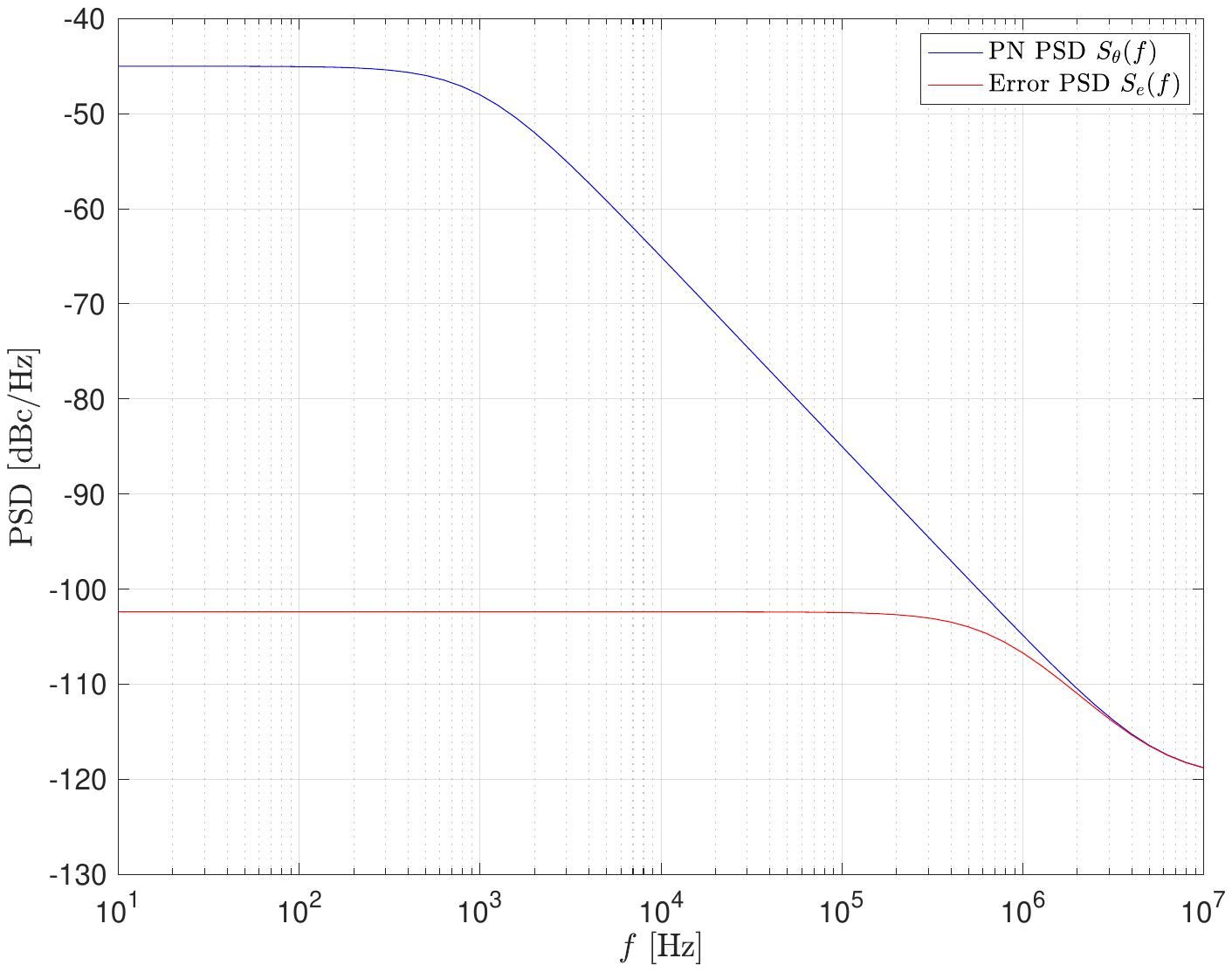}
\caption{Power spectral densities of the phase noise and of the residual phase error after tracking in the system bandwidth.}\label{f:psd_pn_error}
\end{figure}

The PSD of the residual PN $e_k=\theta_k-\hat{\theta}_k$ in the bandwidth $[-1/2\Ts,1/2\Ts ]$ can be obtained by 
using the expression of the PSD of the noise $n'_k$ given in (\ref{e:PSDnoise}) and the expression of the PSD of the PN $\theta^c_k$, given in~(\ref{e:psd_l100}) with $\lii=0$. 
We obtain
\begin{equation}
S_e(f)=\frac{ 10^{10} \lcc  }{  \frac{2 \cdot  10^{10} \lcc}{2\lii+\beta \sigma_w^2\Ts} +\fbb+f^2} + \lii\,.
\end{equation}
From the above expression, we observe that the residual error PSD has the same shape of the PSD of the PN that we wish to estimate, with a 3dB bandwidth $$
\fbe=\sqrt{\fbb  +\frac{2\cdot 10^{10} \lcc }{ \beta \sigma_w^2\Ts +2\lii } } 
$$ 
larger than the one of the PN PSD. Figure~\ref{f:psd_pn_error} shows the two PSDs in the case where $\text{SNR}=10$~dB, the PN is characterized by $\lcc=-85$~dBc/Hz, $\lii=-120$~dBc/Hz, and $\fb=1$~kHz and the symbol rate is $\Rs=10$~MHz.

The performance of a symbol-by-symbol detector, which ignores the correlation among symbols due to the residual PN, depends only on the first order statistics of the PN after tracking. 
For this reason, in the following we compute the variance of the residual PN error, which can be used to predict the performance of the considered system.
Being the process zero-mean, the residual PN error variance, distorting the communication, is obtained by integrating the error spectrum in the communication bandwidth, that is
\begin{align}\nonumber
&\sigma_e^2=\int_{-\frac{1}{2\Ts} }^\frac{1}{2\Ts} S_e(f)df\\ \label{e:varience_error}  
&=\!\frac{2\cdot10^{10} \lcc}{ \sqrt{ \!\fbb \! +\!\frac{2\cdot 10^{10} \lcc }{ \beta \sigma_w^2 \Ts +2\lii }}}  \text{atan}\Bigg(\!\frac{1}{2\Ts \sqrt{ \fbb +\frac{2\cdot 10^{10} \lcc }{ \beta \sigma_w^2 \Ts +2\lii  }  }}   \!\Bigg)\!\!+\!\!\frac{\lii}{\Ts}\,.
\end{align}
The expression above connects the phase error after tracking with the main PN parameters and with the system bandwidth. In the following, we derive simplified forms in the limiting case of \textit{free-running} oscillator, i.e., $\fb\rightarrow 0$, and study the impact of  $\li$ on the performance.

\subsection{Performance in the case of free-running oscillators}

In the case of \textit{free-running} oscillators, i.e., $\fb\rightarrow 0$, the residual error variance is
\begin{align}\nonumber
\sigma_e^2=&  \sqrt{ 2\!\cdot \!10^{10} \lcc (\beta \sigma_w^2 \Ts +2\lii )} \\ \label{e:var_error_freerunning}&\cdot \text{atan}\Bigg(\!\frac{1}{2\Ts }\sqrt{  \frac{2\cdot 10^{10} \lcc }{ \beta \sigma_w^2 \Ts +2\lii  }  }   \!\Bigg)\!\!+\!\!\frac{\lii}{\Ts}\, .  
\end{align}
By approximating the arctangent with $\pi/2$ and using the fact that $\beta \sigma_w^2 \Ts\gg 2\lii$, the above expression becomes the following
\begin{equation}\label{e:varience_error_approx1}
  \sigma_e^2 \simeq   \pi 10^{5} \lc \sigma_w \sqrt{\frac{ \beta  \Ts}{2 }  }   + \frac{\lii}{\Ts}\,,
\end{equation}
which can be also expressed in terms of the phase noise innovation variance given in~(\ref{e:innovation variance}), that is
\begin{equation}
\sigma_e^2 \simeq \sigma_u \sigma_w \sqrt{\frac{\beta}{ 8}}  + \frac{\lii}{\Ts}\,.
\end{equation}

The above simplified expressions are useful for understanding how the residual PN scales with the system parameters. 
Regarding, for example, the symbol rate $\Rs=1/\Ts$, if we consider that $\li$ is usually very low, this result is in line with the observation that in most cases the performance improves if we increase the symbol rate, since the PN varies more slowly from one symbol to the other and hence it is easier to be tracked. On the other hand, there are cases where we observe the opposite behaviour, and this happens when the flat part of the PSD at high frequencies is not negligible. This means that the performance does not depend monotonically on the symbol rate. Particularly, the performance improves by increasing the symbol rate up to a point where the effect of the white phase noise entering the signal bandwidth dominates the performance. 

Expression~(\ref{e:varience_error_approx1}) can explain this phenomenon.
The error variance in~(\ref{e:varience_error_approx1}) is a concave function in \mbox{$\Rs>0$} and hence its minimum can be computed by setting to zero the partial derivative with respect to $\Rs$. 
In this way it is possible to find that the minimum of error variance is achieved in the symbol rate $\Rs^*$
\begin{equation}\label{e:Rmin}
  \Rs^*=\Big( \frac{\pi^2 10^{10} \lcc \beta \sigma_w^2}{8 \li^4} \Big)^{1/3}.  
  \end{equation}
In Section~\ref{s:results}, we will see that this result, although derived under the hypothesis of free-running oscillators, is useful also in the generic case of PLL-locked oscillators.


\section{RESULTS}\label{s:results}
\begin{figure}
\centering
\includegraphics[width=\columnwidth]{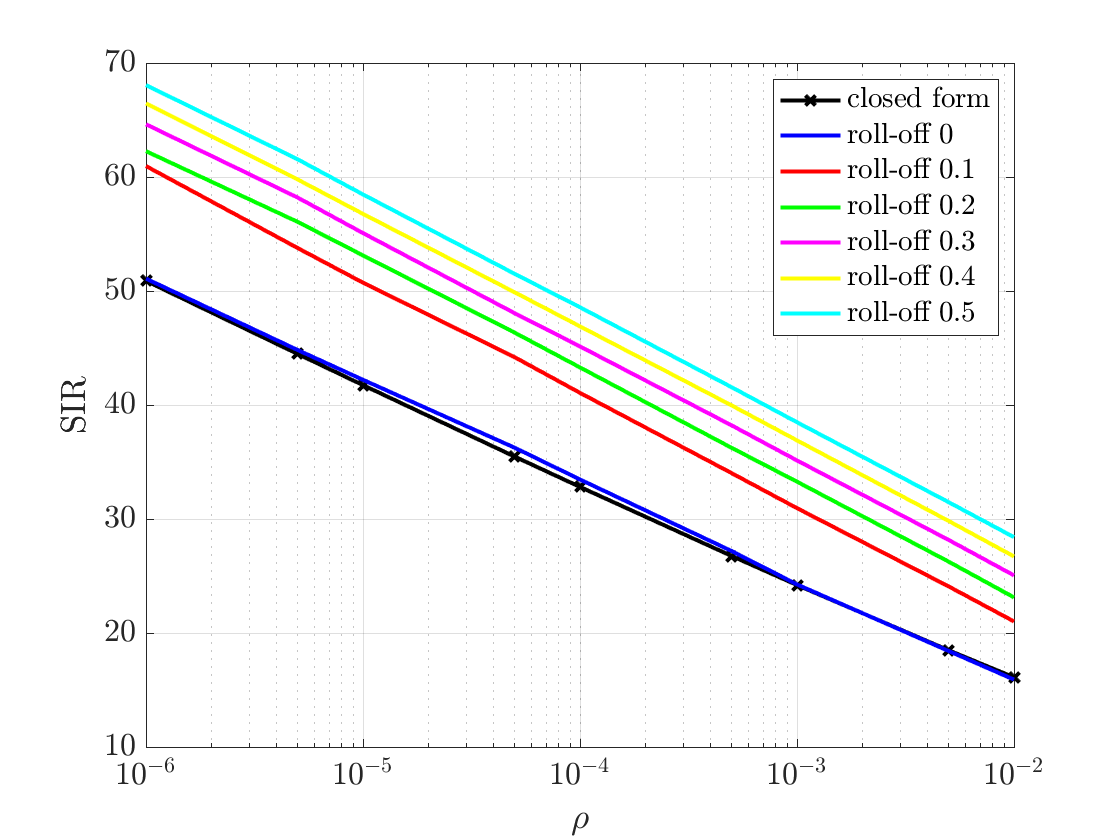}
\caption{SIR after the matched filter at the receiver as a function of the relative bandwidth parameter $\bho$.
}\label{f:SIR}
\end{figure}

We evaluate the SIR after the matched filter and the downsampling operation of the continuous-time system in~(\ref{e:r}), simulated by using an oversampling factor $5$. We consider the transmission of linearly modulated QPSK symbols with a RRC shaping pulse with roll-off factor ranging from 0 to 0.5. The simulated SIRs are reported as a function of the relative bandwidth parameter in Figure~\ref{f:SIR}. 
In order to validate the proposed analysis, we report the curve that corresponds to the closed-form expression given in~(\ref{e:sir}), that is obtained under the assumption that the filter $p(t)$ is the ideal sinc function (roll-off zero). The numerical results are in line with the closed-form analysis, being the simulated curve with roll-off zero almost overlapped with the theoretical one.
Regarding the curves with roll-off larger than zero, they have the same behaviour and approach the closed form as the roll-off decreases.

\begin{figure}
\centering
\includegraphics[width=\columnwidth]{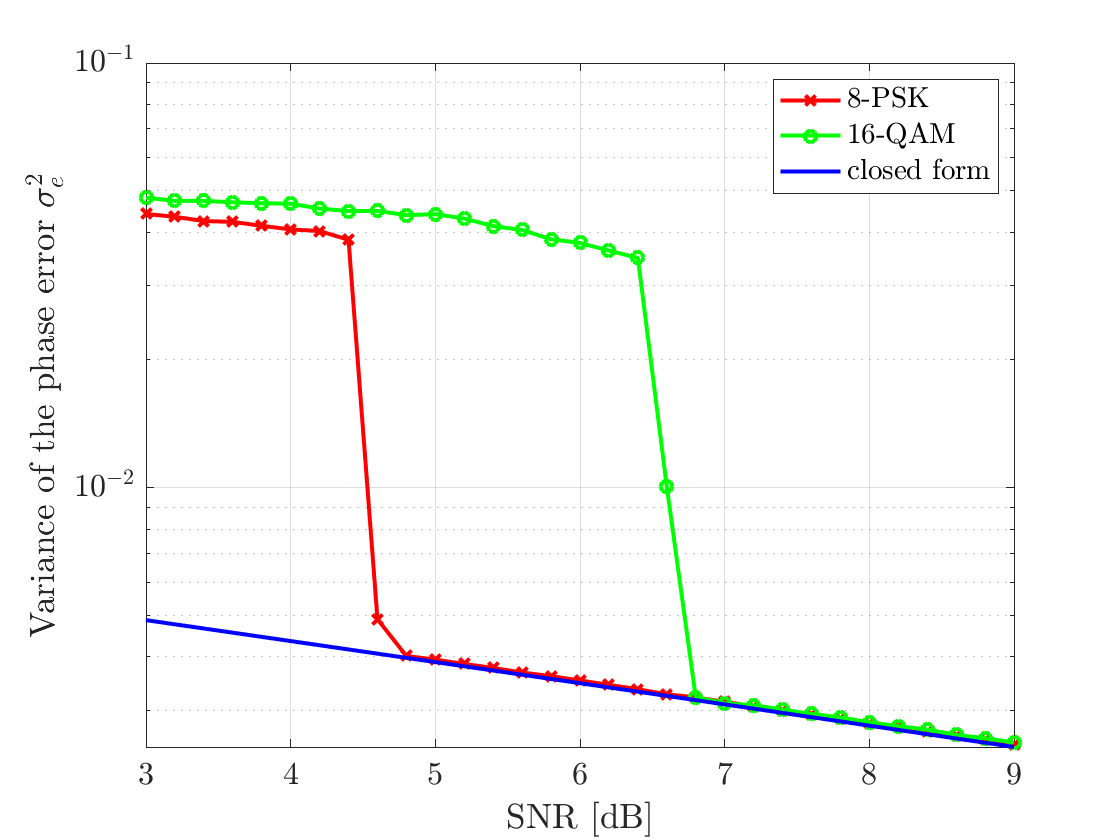}
\caption{Variance of the phase error after tracking. Comparison between the practical Kalman-based estimator and the proposed closed-form expression for $\lcc=-80$.}\label{f:error_m80}
\end{figure}

\begin{figure}
\centering
\includegraphics[width=\columnwidth]{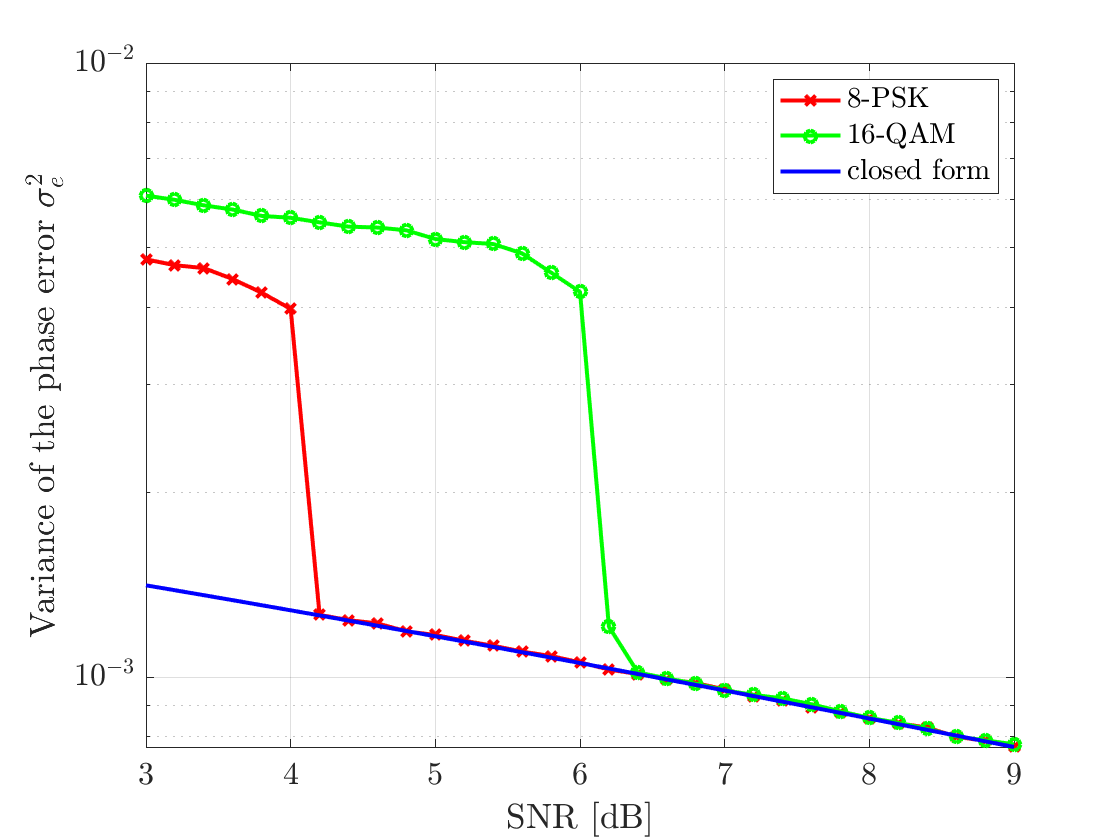}
\caption{Variance of the phase error after tracking. Comparison between the practical Kalman-based estimator and the proposed closed-form expression for $\lcc=-90$.}\label{f:error_m90}
\end{figure}

The analysis about the optimal phase tracking theory is validated by using computer simulations to compute the residual PN error after tracking. We assume to use the low-density parity-check (LDPC) code~\cite{DVB-S2X}, with codeword size 64800 bits and rate $1/2$. The coded bits are mapped into $16$-QAM or $8$-PSK symbols and one pilot symbol is inserted every 50 information symbols. The transmitted symbols are corrupted by PN and AWGN. 
The PN sequence has been generated according to the following parameters: $\Rs=10$~MHz, $\fb=10$~kHz, $\lii=-120$~dBc/Hz and $\lcc=\{-80,-90\}$~dBc/Hz. 
At the receiver, after down conversion, the phase is estimated by using a practical PN estimator, that is a Kalman smoother~\cite{Ka93}. The Kalman estimator is employed in an iterative fashion: at the first iteration, the estimator uses only the pilots symbols, while in the successive ones it uses also the symbol estimates, obtained by running the detection and LDPC decoding algorithms. In Figures~\ref{f:error_m80} and ~\ref{f:error_m90}, we report the variance of the phase error as a function of the SNR for  $\lcc=-80$ and $-90$~dBc/Hz, respectively. The simulation results are compared with the closed-form expression derived in Section~\ref{s:phase_tracking}. The Kalman-based estimator  is run for 3 iterations and, since it uses the knowledge of the modulus of the symbols (exact/approximated in the case of pilots/information symbols), we set $\beta=1$ in (\ref{e:varience_error}). The figure shows that the curves of the practical estimators follow the ones obtained in closed form after a given SNR, that corresponds to the SNR where the estimates fed back from the decoder to the phase tracker are more reliable and hence the Kalman estimator is effective. This means that the hypothesis of known symbols, under which the closed form of the error variance was obtained, is fulfilled with good approximation. It is worth noticing that, for fixed PN parameters, the gap between the SNR where the simulations correspond to the theory for $8$-PSK and $16$-QAM is around $2$ dB. This corresponds to the SNR gap in the Shannon capacity curve between $1.5$ and $2$ [bps/Hz], that are the mutual information of $8$-PSK and $16$-QAM coupled with a binary code with rate $1/2$, respectively.

\begin{figure}
\centering
\includegraphics[width=\columnwidth]{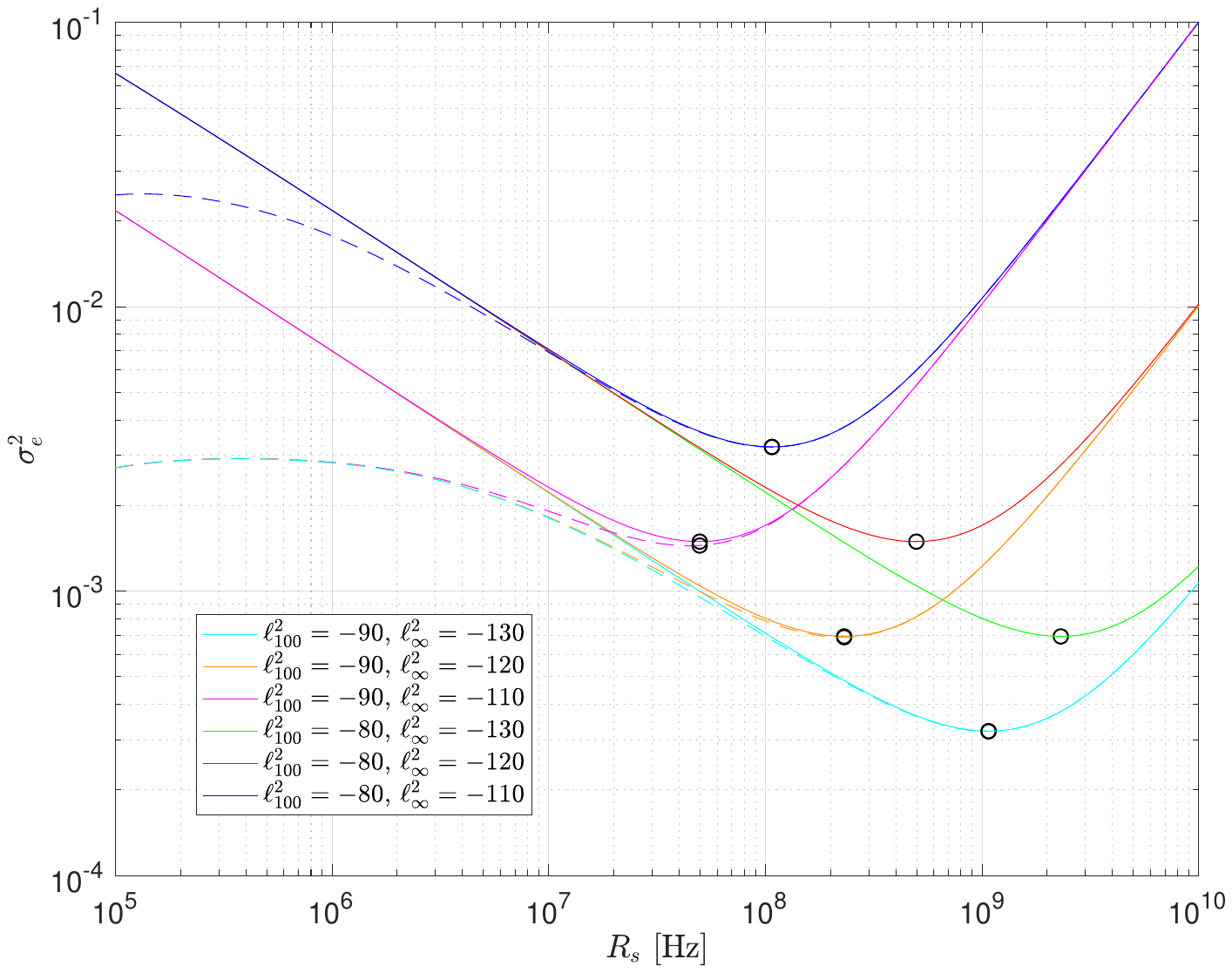}
\caption{Error variance as a function of the symbol rate for PLL-locked (continuous line) and free-running (dashed line) oscillators. The minimum of the variance obtained through~(\ref{e:Rmin}) is indicated by circles.}\label{f:minimum}
\end{figure}

In Figure~\ref{f:minimum}, we consider the variance of the residual error computed through the proposed closed-form expression as a function of the symbol rate $R_s$ for \mbox{$\text{SNR}=0$~dB}, $\lii=\{-130, -120, -110\}$~dBc/Hz and $\lcc=\{-80,-90\}$~dBc/Hz. The dashed line are for the PLL-case, with $\fb=10$~kHz, while the continuous lines are for the free-running oscillator case. The figure shows that the variance decreases up to a point where increasing the symbol rate is not beneficial anymore. On the contrary, the performance degrades due to the presence of the white PN. We observe that this change of trend happens for lower symbol rate when $\lii$ is larger. In the figure we report with a circle the argmin of the error variance computed through~(\ref{e:Rmin}), and show that this computation is accurate and valid for both free-running and PLL-locked oscillators.

\section{CONCLUSION}
In this paper, the phase noise typical of local oscillators in communication systems has been considered. The discrete-time phase noise channel has been studied by using analytical models described by means of parameters linked to measurements.
The intersymbol interference and the power loss due to the presence of phase noise, affecting the samples obtained by matched filtering and sampling, have been bounded. 

Then, we have studied the performance of phase tracking algorithms. We have applied the optimal estimation theory to find an expression for the residual error variance after tracking as a function of the PN parameters, the AWGN variance and the system bandwidth. Moreover, we derived in closed form the expression for the optimal symbol rate that minimizes the estimation error variance. Simulation results based on a practical Kalman based estimator have validated the proposed theory.

An important implication of our work is that the derived closed-form expressions can be used in the design of receiver algorithms and for performance prediction, starting from the knowledge of some fundamental parameters that can be found from measurements of practical oscillators.


\section*{ACKNOWLEDGMENT}

\bibliographystyle{IEEEtran}

\end{document}